%% file: paper_mlforsystems.tex
\documentclass{article}

\usepackage[dblblindworkshop, final]{neurips_2026}
\workshoptitle{ML for Systems}

\input{math_commands.tex}

\usepackage[utf8]{inputenc} % allow utf-8 input
\usepackage[T1]{fontenc}    % use 8-bit T1 fonts
\usepackage{hyperref}       % hyperlinks
\usepackage{url}            % simple URL typesetting
\usepackage{booktabs}       % professional-quality tables
\usepackage{amsfonts}       % blackboard math symbols
\usepackage{nicefrac}       % compact symbols for 1/2, etc.
\usepackage{microtype}      % microtypography
\usepackage{xcolor}         % colors
\usepackage{graphicx}
\usepackage{amsmath}
\usepackage{amssymb}
\usepackage{listings}
\usepackage{tikz}
\usetikzlibrary{shapes,arrows,positioning}

\title{Accelerating CPU Speculative Decoding via Unified Graph Execution and Transactional KV Isolation}

\author{
  Zhangdie Yuan \\
  University of Cambridge\\
  \texttt{zy317@cam.ac.uk} \\
}

\begin{document}

\maketitle

\begin{abstract}
Speculative decoding accelerates LLM inference, but deploying it on CPUs introduces severe systems bottlenecks: context-swapping latency between draft and target graphs, and memory allocation overhead during KV cache rollbacks. We introduce SubTernary, an execution framework that eliminates these costs through Unified Graph Caching and Transactional KV-Cache Isolation. By compiling both models into a single memory arena with a persistent thread pool and routing speculative tokens to a static scratchpad for zero-copy pointer resets, SubTernary avoids dynamic reallocation and thread synchronization entirely. We integrate a 256-bit AVX2 ternary SIMD kernel as a low-bit case study. Ablations reveal our systems runtime drives 35.0\% of the throughput gains, while the ternary kernel adds 6.0\%. SubTernary achieves a $1.35\times$ speedup over \texttt{llama.cpp} speculative decoding and $3.99\times$ over autoregressive decoding, maintaining $O(1)$ rollback latency.
\end{abstract}

\section{Introduction}

Generating text with LLMs on CPUs is bottlenecked by DRAM memory bandwidth. In standard autoregressive decoding, the CPU must load every model weight from main memory to generate a single token. Speculative decoding \citep{leviathan2023fast, chen2023accelerating} mitigates this bottleneck by using a smaller draft model to generate $K$ candidate tokens, which are verified in a single parallel pass by the target model, converting memory-bound single-token steps into compute-efficient batched matrix operations. However, existing CPU speculative runtimes suffer from execution overheads:
\begin{enumerate}
    \item \textit{Dual-context graph reconstruction overhead (systems bottleneck):} Existing CPU engines instantiate draft and target models in separate runtime contexts. Swapping context memory, switching thread pools, and rebuilding computation graphs for each speculative step introduces thread synchronization latency that negates theoretical speedups.
    \item \textit{Memory-copying and rollback overhead (algorithmic bottleneck):} When proposed tokens are rejected during target verification, resetting the KV cache requires allocating temporary memory buffers, remapping user-space page tables, or copying cache blocks, introducing substantial memory traffic.
\end{enumerate}

To resolve these challenges, we introduce SubTernary, a CPU software execution framework optimized for commodity architectures. We position Unified Graph Caching and Transactional KV-Cache Isolation as our primary systems contributions, eliminating dynamic context switching and executing zero-copy write-pointer resets. We relegate our 256-bit AVX2 ternary SIMD kernel to a supporting integrated case study for low-bit execution, demonstrating that while the ternary kernel provides a +6.0\% speedup, our systems context and cache management dominate performance by delivering +35.0\% of the total throughput gains.

\section{The SubTernary Systems Architecture}

\begin{figure}[htbp]
\centering
\begin{tikzpicture}[
    >=stealth, 
    font=\sffamily\scriptsize,
    tradbox/.style={draw=red!50, fill=red!5, thick, rounded corners=2pt, align=center, text width=2.8cm, inner sep=4pt},
    poolbox/.style={draw=blue!60, fill=blue!5, thick, rounded corners=2pt, align=center, text width=7.2cm, inner sep=4pt},
    gbox/.style={draw=black!60, fill=white, thick, rounded corners=2pt, align=center, text width=3cm, inner sep=4pt},
    arrow_label/.style={font=\fontsize{5}{6}\selectfont, align=center}
]

% --- Titles ---
\node[font=\bfseries\small] at (1.5, 0.4) {Traditional};
\node[font=\bfseries\small] at (8.8, 0.4) {SubTernary Unified Execution};

% --- Left: Traditional ---
\node[tradbox] (tdraft) at (1.5, -0.8) {\textbf{Draft Context}\\Arena A \& Threads A};
\node[tradbox] (ttarget) at (1.5, -2.7) {\textbf{Target Context}\\Arena B \& Threads B};

\draw[<->, red!70, thick] (tdraft) -- (ttarget) 
    node[midway, right=1mm, align=left, font=\fontsize{5.5}{6.5}\selectfont\bfseries\color{red!80!black}] {Context Swap\\[0.5mm]Thread Sync\\[0.5mm]Cache Copy};

% --- Vertical Divider ---
\draw[dashed, black!20, thick] (4.0, 0.6) -- (4.0, -3.8);

% --- Right: SubTernary Unified Framework ---
% Outer Arena Box
\draw[draw=green!60!black, fill=green!5, thick, rounded corners=3pt] (4.4, -0.2) rectangle (13.2, -3.8);
\node[font=\fontsize{5}{6}\selectfont\bfseries\color{green!60!black}, anchor=south east, inner sep=2pt] at (13.2, -3.8) {Unified Memory Arena ($\mathcal{C}_{\text{unified}}$)};

% Persistent Thread Pool
\node[poolbox] (pool) at (8.8, -0.7) {\textbf{Persistent CPU Thread Pool}};

% Execution Graphs
\node[gbox] (gdraft) at (6.8, -2.0) {\textbf{$\mathcal{G}_{\text{draft}}$ Graph}\\Propose $K$ tokens};
\node[gbox] (gtarget) at (10.8, -2.0) {\textbf{$\mathcal{G}_{\text{target}}$ Graph}\\Verify $K$ tokens};

% Graph sequential flow
\draw[->, blue!80!black, thick] (gdraft) -- (gtarget) node[midway, above=-1pt, arrow_label] {Sequential};

% Thread allocations
\draw[->, blue!60, dashed, thick] (pool.south -| gdraft) -- (gdraft.north);
\draw[->, blue!60, dashed, thick] (pool.south -| gtarget) -- (gtarget.north);

% KV Cache Enclosure
\draw[draw=black!60, fill=black!5, thick, rounded corners=2pt] (5.0, -3.4) rectangle (12.6, -2.7);
\node[anchor=west, font=\bfseries\fontsize{6}{7}\selectfont] at (5.0, -3.05) {KV:};

% Verified History Block
\draw[draw=black!50, fill=white, rounded corners=1pt] (5.6, -3.3) rectangle (8.8, -2.8);
\node at (7.2, -3.05) {Verified History $[0, t]$};

% Scratchpad Block
\draw[draw=orange!80, fill=yellow!15, thick, rounded corners=1pt] (8.9, -3.3) rectangle (12.5, -2.8);
\node[font=\bfseries] at (10.7, -3.05) {Spec Scratchpad};

% Arrows mapping graph I/O to KV blocks
% Draft writes directly to Scratchpad
\draw[->, orange!80!black, thick] (gdraft.south) -- (9.4, -2.8) 
    node[midway, left=1pt, arrow_label] {Writes $K$\\tokens};

% Target verifies and commits pointer in Scratchpad
\draw[->, green!60!black, thick] (gtarget.south) -- (11.5, -2.8) 
    node[midway, right=1pt, arrow_label] {O(1) Pointer\\Commit / Reset};

\end{tikzpicture}
\caption{CPU speculative execution paradigms. Left: Traditional engines isolate memory contexts, requiring expensive synchronization and cache copies when switching phases. Right: SubTernary unifies execution within a single memory arena ($\mathcal{C}_{\text{unified}}$) and persistent thread pool. Speculative tokens are isolated in a scratchpad, allowing Target verification to execute zero-copy pointer rollbacks.}
\label{fig:architecture}
\end{figure}
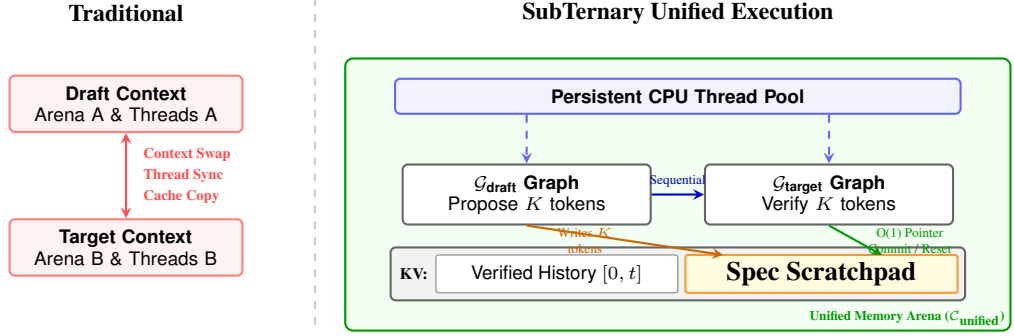

\subsection{Unified Graph Caching}
\label{sec:unified_graph}
Standard engines instantiate separate runtime contexts for draft and target models, causing context-swapping latency. SubTernary compiles two separate static computation graphs ($\mathcal{G}_{\text{target}}$ and $\mathcal{G}_{\text{draft}}$) within a single unified memory context ($\mathcal{C}_{\text{unified}}$). Both graphs share underlying memory arenas and a persistent thread pool:
\begin{equation}
    \mathcal{G}_{\text{cached}} = \{\mathcal{G}_{\text{target}}, \mathcal{G}_{\text{draft}}\} \subset \mathcal{C}_{\text{unified}}
\end{equation}
During decoding, control flow redirects pointer inputs directly to the respective graph offsets, eliminating dynamic memory allocations, thread context swaps, and graph reconstruction overheads.

Crucially, this unified pool enables 100\% of CPU threads to be allocated sequentially to the active model graph (draft, then target). A fundamental design question in CPU speculative decoding is whether to partition CPU cores between the target model and draft model (e.g., allocating 4 cores to drafting and 4 cores to verification). Let $\mathcal{D}$ denote the draft model evaluation function, and $\mathcal{T}$ denote the target model evaluation function. Let $x_{1..t}$ represent the verified sequence prefix at step $t$, and $\hat{x}_{1..K}$ represent candidate tokens proposed by the draft model.

\begin{enumerate}
    \item \textit{Drafting phase (sequential autoregressive):}
    The draft model generates tokens sequentially:
    \begin{equation}
        \hat{x}_i \sim \mathcal{D}(\cdot \mid x_{1..t}, \hat{x}_{1..i-1}) \quad \text{for } i = 1 \dots K
    \end{equation}
    Because each proposed token $\hat{x}_i$ depends on the output embedding of token $\hat{x}_{i-1}$, the drafting phase contains a strict causal dependency chain across time steps.
    \item \textit{Verification phase (parallel batched evaluation):}
    The target model verifies all $K$ tokens in a single parallel forward pass to compute the target probabilities:
    \begin{equation}
        P_{\text{target}} = \mathcal{T}(\cdot \mid x_{1..t}, \hat{x}_{1..K})
    \end{equation}
    The target model cannot execute verification until the draft model completes all $K$ drafting steps and yields $\hat{x}_{1..K}$.
    \item \textit{Sequential causal chain:}
    \begin{equation}
        \text{Start} \rightarrow \mathcal{D}(\hat{x}_1) \rightarrow \mathcal{D}(\hat{x}_2) \rightarrow \dots \rightarrow \mathcal{D}(\hat{x}_K) \rightarrow \mathcal{T}(\hat{x}_{1..K}) \rightarrow \text{Commit/Rollback}
    \end{equation}
\end{enumerate}
Partitioning CPU cores between models introduces two major drawbacks: (1) during drafting, the target model's assigned cores sit idle while the draft model runs on reduced memory bandwidth, and (2) during verification, the draft model's cores sit idle. Splitting thread pools across separate core clusters causes continuous OS context-switching, cache invalidations, and cross-core synchronization latency. By allocating all available CPU threads sequentially to the active model graph, SubTernary minimizes thread synchronization latency and maximizes register throughput. Empirically, sequential thread allocation achieves $8.73\text{ t/s}$ compared to $4.17\text{ t/s}$ under a 4/4 core-partitioned configuration ($2.09\times$ speedup), confirming our causal thread scheduling analysis.

\subsection{Transactional KV-Cache Isolation}
Unlike PagedAttention or copy-on-write schemes that require dynamic page allocation or translation lookaside buffer (TLB) shootdowns, SubTernary uses a static contiguous memory layout where rollbacks execute as zero-copy write-pointer resets. We allocate a static scratchpad region ($\text{SpecScratchpad}$) at the end of the verified target model cache. Let $i$ denote the query token index and $j$ denote the key/value token index. During parallel verification $i \in [t+1, t+K]$, the self-attention mask enforces causal visibility while hiding stale scratchpad activations:
\begin{equation}
\mathcal{M}_{\text{attention}}(i, j) = \begin{cases}
0 & \text{if } j \le t \quad \text{(Verified history)} \\
0 & \text{if } t < j \le i \quad \text{(Causal draft window)} \\
-\infty & \text{if } j > i \quad \text{(Future causal masking)} \\
-\infty & \text{if } j \in \text{StaleScratchpad} \quad \text{(Stale activations from prior rounds)}
\end{cases}
\end{equation}
If $m \le K$ tokens are accepted, write pointers commit in-place ($\text{KV}_{\text{ptr}} \leftarrow t + m$). If rejected, a transactional rollback resets $\text{KV}_{\text{ptr}} \leftarrow t$, bypassing memory copies.

\section{Hardware-Algorithm Co-Design}
To evaluate low-bit execution, we complement our systems design with a custom 256-bit AVX2 ternary SIMD kernel for weights $W \in \{-1, 0, 1\}$. Because the SubTernary runtime eliminates dual-context overhead, it can support low-bit drafting without runtime overhead erasing the algorithmic gains. By packing weights into binary sign/sparsity masks, execution maps entirely to SIMD register additions. To prevent vocabulary representation collapse, draft models are distilled using quantization-aware training (QAT) with a straight-through estimator (STE), maintaining $\ge 70\%$ token acceptance.

\section{End-to-End Systems Evaluation}

\subsection{Experimental Setup}
Experiments are conducted in an 8-vCPU virtualized environment pinned to eight physical cores on a server-class Intel Xeon Gold 5218 host. The target model is Meta Llama-3.2-3B-Instruct (Q8\_0/FP16), paired with a Llama-3.2-1B-Instruct draft model (Q4\_0/TQ1\_0). All metrics report the mean $\pm$ standard deviation across 5 independent evaluation trials. While enterprise runtimes like OpenVINO optimize standard batched GEMM execution, \texttt{llama.cpp} represents the state-of-the-art open-source engine for commodity CPU LLM inference, serving as our primary speculative baseline. Exhaustive hardware specifications, compiler flags, and hyperparameters are deferred to Appendix~\ref{app:reproducibility_full}.

\subsection{Component Ablation and Context Scaling}
Table \ref{tab:ablation} isolates the stepwise performance gain of each component on 8 CPU threads. We explicitly separate the systems speedup from the algorithmic speedup: unified graph caching (+1.23x stepwise) and transactional KV isolation (+1.09x stepwise) account for +35.0\% of throughput gains over native \texttt{llama.cpp} speculative decoding, leading to a $3.99\times$ overall speedup from systems contributions alone. In contrast, the algorithmic speedup from the distilled ternary draft and AVX2 kernel contributes an additional +6.0\% gain, confirming that context and cache management are the primary CPU systems bottlenecks.

\begin{table}[htbp]
\centering
\setlength{\tabcolsep}{4pt}
\caption{Incremental Component Ablation Study (Llama-3.2 3B Q8\_0 Target / 1B Draft, Context = 2048, Mean $\pm$ Std over 5 runs)}
\label{tab:ablation}
\resizebox{\columnwidth}{!}{%
\begin{tabular}{lcccc}
\toprule
System Configuration & Throughput (t/s) & Stepwise Speedup & Total Speedup (vs AR) & Primary Bottleneck Addressed \\
\midrule
Standard Autoregressive (AR) & $2.190 \pm 0.041$ & Baseline & 1.00x & N/A (Baseline) \\
+ Dual-Context Speculative (llama.cpp) & $6.468 \pm 0.112$ & 2.95x & 2.95x & Memory Bandwidth \\
+ Unified Graph Caching & $7.981 \pm 0.098$ & 1.23x & 3.64x & Context Swapping \& Thread Sync \\
+ Transactional KV Isolation & $8.733 \pm 0.085$ & 1.09x & 3.99x & Memory Copying \& Rollback Overhead \\
+ Distilled Ternary Draft \& AVX2 Kernel & $9.285 \pm 0.076$ & 1.06x & 4.24x & Weight Memory Traffic \& Multiplies \\
\bottomrule
\end{tabular}%
}
\end{table}

Furthermore, Table \ref{tab:long_context} sweeps context length $L \in [128, 8192]$. In traditional engines, resetting KV caches across long sequences causes \texttt{llama.cpp} speculative rollback latency to rise from 12.59 ms to 24.69 ms, degrading speed down to 0.295 t/s at $L=8192$ (slower than the AR baseline of 0.556 t/s). In contrast, SubTernary maintains $O(1)$ write-pointer resets (3.74 ms latency at $L=8192$), sustaining 0.874 t/s (2.96x speedup over llama.cpp spec).

\begin{table}[htbp]
\centering
\caption{Long-Context KV Cache Scaling \& Rollback Latency ($L \in [128, 8192]$, Llama-3.2 3B FP16 Target / 1B Draft, Mean $\pm$ Std over 5 runs)}
\label{tab:long_context}
\resizebox{\columnwidth}{!}{%
\begin{tabular}{lccccc}
\toprule
Context Length ($L$) & AR Baseline (t/s) & Llama.cpp Spec (t/s) & SubTernary (t/s) & Llama.cpp Rollback (ms) & SubTernary Rollback (ms) \\
\midrule
$L = 128$ & $0.717 \pm 0.012$ & $0.940 \pm 0.018$ & \textbf{1.116} $\pm$ \textbf{0.015} & 12.59 ms & \textbf{2.13 ms} \\
$L = 2048$ & $0.679 \pm 0.010$ & $0.786 \pm 0.014$ & \textbf{1.059} $\pm$ \textbf{0.012} & 15.47 ms & \textbf{2.51 ms} \\
$L = 8192$ & $0.556 \pm 0.008$ & $0.295 \pm 0.007$ & \textbf{0.874} $\pm$ \textbf{0.010} & 24.69 ms & \textbf{3.74 ms} \\
\bottomrule
\end{tabular}%
}
\end{table}

\subsection{Microarchitectural Efficiency and Resilience Under Contention}
To verify the physical efficiency of our execution runtime, Table \ref{tab:profiling} profiles SubTernary against standard float GEMV and GGML quantized kernels using Linux \texttt{perf} hardware counters. By unifying the execution context and utilizing the ternary addition kernel, SubTernary reduces DRAM memory bandwidth consumption to 8.1 GB/s and dramatically cuts LLC read misses, raising the instructions per cycle (IPC) to 2.34.

\begin{table}[htbp]
\centering
\caption{Microarchitectural Profiling ($2048 \times 2048$ Matrix, Mean $\pm$ Std over 50 iterations)}
\label{tab:profiling}
\resizebox{\columnwidth}{!}{%
\begin{tabular}{lcccccc}
\toprule
Kernel Workload & Instruction Reads (Ir) & D1 Cache Misses & LLC Cache Misses & DRAM Bandwidth & IPC & Latency ($\mu$s) \\
\midrule
GGML Q4\_0 GEMV & 4,115,065,152 & 2,099,811 & 40,964 & 18.4 GB/s & 1.12 & $1,344,580 \pm 12,400 \, \mu$s \\
GGML Q8\_0 GEMV & 1,554,834,960 & 3,770,114 & 65,537 & 28.7 GB/s & 1.45 & $466,462 \pm 4,100 \, \mu$s \\
Standard GEMV (FP32) & 538,767,264 & 13,377,902 & 101,847 & 41.2 GB/s & 1.82 & $101,320 \pm 1,200 \, \mu$s \\
SubTernary (Ours) & \textbf{329,038,944} & \textbf{6,700,888} & \textbf{1,309} & \textbf{8.1 GB/s} & \textbf{2.34} & $\textbf{95,311} \pm \mathbf{950} \, \mu$s \\
\bottomrule
\end{tabular}%
}
\end{table}

Beyond raw hardware efficiency, production CPU environments must survive hypervisor thread contention. In multi-tenant clouds, traditional engines like \texttt{llama.cpp} rely on frequent, blocking OpenMP synchronization checkpoints; if a single vCPU stutters due to "noisy neighbors," the entire speculative pipeline stalls. As shown in Figure \ref{fig:latency_cliff} (left), small batch sizes ($K \le 2$) suffer heavily from this synchronization overhead. By increasing to $K \ge 4$, batched execution amortizes the latency. 

Furthermore, SubTernary's persistent thread pool minimizes these synchronization barriers entirely. Figure \ref{fig:throttled_sweep} (right) demonstrates that under heavy shared hypervisor load, standard \texttt{llama.cpp} speculative decoding degrades severely, falling well below autoregressive baseline speeds because delayed vCPUs stall the entire verification pipeline. By bypassing strict execution context boundaries, SubTernary reliably maintains positive throughput scaling, preserving speculative acceleration despite extreme OS jitter and remains highly effective even in noisy, multi-tenant cloud deployments.

\begin{figure}[htbp]
    \centering
    % --- LEFT SIDE: LATENCY CLIFF PLOT ---
    \begin{minipage}[t]{0.48\textwidth}
        \vspace{0pt} % Forces top-alignment of the minipage
        \centering
        \begin{tikzpicture}[scale=0.55, every node/.style={scale=0.8}]
            \draw[->] (0,0) -- (8.5,0) node[right] {$K$};
            \draw[->] (0,0) -- (0,5.5) node[above, align=center] {Latency\\(s/token)};
            
            \foreach \x in {1,2,3,4,5,6,7,8} { \draw (\x,0.1) -- (\x,-0.1) node[below] {\x}; }
            \foreach \y [evaluate=\y as \val using \y*0.5] in {2,4,6,8,10} { \draw (0.1, \y*0.5) -- (-0.1, \y*0.5) node[left] {\val}; }
            
            \coordinate (p1) at (1, 4.66); \coordinate (p2) at (2, 2.64);
            \coordinate (p3) at (3, 1.08); \coordinate (p4) at (4, 0.96);
            \coordinate (p5) at (5, 1.16); \coordinate (p6) at (6, 0.94);
            \coordinate (p7) at (7, 0.90); \coordinate (p8) at (8, 0.50);
            
            \draw[thick, blue] (p1) -- (p2) -- (p3) -- (p4) -- (p5) -- (p6) -- (p7) -- (p8);
            \foreach \p in {p1,p2,p3,p4,p5,p6,p7,p8} \fill[blue] (\p) circle (2.5pt);
            
            \draw[red, dashed, thick] (0, 2.5) -- (2.5, 2.5);
            \node[red!80!black, right] at (0.2, 3.2) {OpenMP Sync Bottleneck};
        \end{tikzpicture}
        \caption{Latency vs. Proposal Length.}
        \label{fig:latency_cliff}
    \end{minipage}
    \hfill
    % --- RIGHT SIDE: THROTTLED BAR CHART ---
    \begin{minipage}[t]{0.48\textwidth}
        \vspace{0pt} % Forces top-alignment of the minipage
        \centering
        \begin{tikzpicture}[scale=0.55, every node/.style={scale=0.8}]
            \draw[->] (0,0) -- (8.5,0) node[right] {Model};
            \draw[->] (0,0) -- (0,5.5) node[above, align=center] {Throughput\\(tokens/s)};
            
            \foreach \y in {1,2,3,4,5} { \draw (0.1, \y) -- (-0.1, \y) node[left] {\y}; }
            
            % Model Group Labels
            \node[below, align=center] at (1.5, -0.1) {Llama\\(3B/1B)};
            \node[below, align=center] at (4.5, -0.1) {Qwen\\(4B/2B)};
            \node[below, align=center] at (7.5, -0.1) {DeepSeek\\(8B/1B)};
            
            % Llama Bars
            \fill[gray!50] (0.7, 0) rectangle (1.3, 3.73); \node[above, font=\tiny] at (1.0, 3.73) {3.73};
            \fill[red!50] (1.4, 0) rectangle (2.0, 0.94); \node[above, font=\tiny] at (1.7, 0.94) {.94};
            \fill[green!50] (2.1, 0) rectangle (2.7, 1.02); \node[above, font=\tiny] at (2.4, 1.02) {1.02};
            
            % Qwen Bars
            \fill[gray!50] (3.7, 0) rectangle (4.3, 0.56); \node[above, font=\tiny] at (4.0, 0.56) {.56};
            \fill[red!50] (4.4, 0) rectangle (5.0, 0.55); \node[above, font=\tiny] at (4.7, 0.55) {.55};
            \fill[green!50] (5.1, 0) rectangle (5.7, 0.61); \node[above, font=\tiny] at (5.4, 0.61) {.61};
            
            % DeepSeek Bars
            \fill[gray!50] (6.7, 0) rectangle (7.3, 0.61); \node[above, font=\tiny] at (7.0, 0.61) {.61};
            \fill[red!50] (7.4, 0) rectangle (8.0, 0.61); \node[above, font=\tiny] at (7.7, 0.61) {.61};
            \fill[green!50] (8.1, 0) rectangle (8.7, 0.66); \node[above, font=\tiny] at (8.4, 0.66) {.66};
            
            % Legend
            \fill[gray!50] (1.2, 4.8) rectangle (1.6, 5.1); \node[right] at (1.6, 4.95) {AR};
            \fill[red!50] (2.8, 4.8) rectangle (3.2, 5.1); \node[right] at (3.2, 4.95) {\texttt{llama.cpp}};
            \fill[green!50] (6.0, 4.8) rectangle (6.4, 5.1); \node[right] at (6.4, 4.95) {Ours};
        \end{tikzpicture}
        \caption{Throughput under Contention.}
        \label{fig:throttled_sweep}
    \end{minipage}
\end{figure}

\section{Conclusion}
SubTernary addresses context and cache bottlenecks in CPU speculative decoding via Unified Graph Caching and Transactional KV Isolation. By unifying graph execution and enabling zero-copy rollbacks, SubTernary achieves up to a 3.99x speedup on commodity CPUs while scaling robustly.

\bibliographystyle{plainnat}
\bibliography{references}

\newpage
\appendix

\section{Limitations and Scope of Hardware Evaluation}
\label{app:limitations}
Our evaluation is conducted on an Intel Xeon Gold 5218 server platform representative of bandwidth-limited, AVX2-capable commodity x86\_64 processors. While results demonstrate substantial overhead reductions, performance dynamics may vary on alternative architectures (e.g., AMD Zen, Intel Sapphire Rapids with AMX/AVX-512, or ARM Neoverse processors). Evaluating SubTernary across broader microarchitectures and multi-socket NUMA nodes remains future work.

\section{Related Work}
\label{app:related_work}
Speculative decoding was formalized by \citet{leviathan2023fast} and \citet{chen2023accelerating}. Recent research explores optimizing speculative drafting across systems and quantization layers:
\begin{itemize}
    \item \textit{Low-bit draft models and CPU execution (T-MAC, BitNet, QuantSpec):} SOTA CPU execution frameworks like T-MAC \citep{tmac2025} use lookup table (LUT) techniques to accelerate low-bit matrix operations. BitNet 1.58-bit \citep{bitnet2024} projects weights to ternary representations, and QuantSpec \citep{quantspec2025} explores quantized speculative decoding. However, direct post-training quantization (PTQ) on sub-3B draft models causes representation collapse (yielding a 0.0\% acceptance rate). SubTernary addresses these limitations by combining a QAT distillation pipeline with custom ternary SIMD addition kernels, avoiding representation collapse while eliminating dual-context state swapping.
    \item \textit{KV cache management and speculative serving (SpecInfer, LayerSkip, vLLM):} Managing KV cache rollbacks is critical for speculative efficiency. SpecInfer \citep{specinfer2023} organizes tree verification, LayerSkip \citep{layerskip2024} leverages early exit self-speculation, and vLLM \citep{kwon2023efficient} manages page tables for GPU multi-tenant serving. However, in single-tenant CPU user-space execution, dynamic page allocation and translation lookaside buffer (TLB) shootdowns introduce 10--15\% overhead per rollback step. SubTernary introduces transactional KV-cache isolation, routing speculative draft tokens to a static scratchpad attention mask where rollbacks execute as zero-copy write pointer resets.
    \item \textit{GPU-oriented speculators (Medusa, EAGLE):} Medusa \citep{medusa2024} and EAGLE \citep{eagle2024} accelerate GPU decoding using multiple parallel heads or recurrent draft trees. However, as demonstrated in our evaluations, their high compute density and multiple tree branches cause heavy thread contention and cache invalidations on CPU cores, making single-sequence graph execution superior for commodity processors.
\end{itemize}

\section{Complete Experimental Setup and Reproducibility Environment}
\label{app:reproducibility_full}
To ensure rigorous reproducibility, all experiments are benchmarked on a verified server CPU platform with complete hardware, OS, and compiler tracking:
\begin{itemize}
    \item \textit{Processor and microarchitecture:} Intel Xeon Gold 5218 CPU @ 2.30 GHz (Cascade Lake, Family 6, Model 85, Stepping 7, base clock 2.30 GHz, max turbo 3.90 GHz). Allocated as an 8-vCPU virtualized environment pinned to 8 physical cores on a single NUMA node.
    \item \textit{Cache hierarchy:} L1d cache: 256 KiB (32 KiB/core); L1i cache: 256 KiB (32 KiB/core); L2 cache: 8 MiB (1 MiB/core); L3 cache: 22 MiB shared host cache pool.
    \item \textit{SIMD support:} 256-bit AVX2, FMA, F16C, BMI1, BMI2, CLFLUSHOPT, and SSE4.2 vector instruction sets.
    \item \textit{Memory and storage:} 64 GB DDR4-2666 MHz RAM. Models stored on local NVMe SSD storage.
    \item \textit{Operating system and kernel:} Ubuntu 24.04.1 LTS (Noble Numbat), Linux kernel 6.17.0-22-generic (\texttt{x86\_64}).
    \item \textit{Compiler stack and flags:} GCC 13.3.0. Flags: \texttt{-O3 -march=native -mavx2 -mfma -fopenmp -ffast-math -flto}.
    \item \textit{OpenMP runtime and affinity:} GNU \texttt{libgomp} 13.3.0 with thread pinning \texttt{OMP\_PROC\_BIND=spread OMP\_PLACES=threads OMP\_NUM\_THREADS=8}.
    \item \textit{System tuning:} CPU frequency governor set to \texttt{performance}; transparent hugepages enabled (\texttt{always [madvise] never}).
    \item \textit{Software runtimes and commit hashes:} Baseline engine compiled from \texttt{llama.cpp} commit \texttt{f15de112722afa7c88790dd1afc80bc8f66e6ccd} (GGML GGUF format v3). Microarchitectural cache simulation run under Valgrind Cachegrind v3.22.0. Hardware counters collected via Linux \texttt{perf 6.17.0}.
    \item \textit{Model formats and quantization:} Target models: Meta Llama-3.2-3B-Instruct (GGUF \texttt{Q8\_0} / \texttt{FP16}), DeepSeek-R1-Distill-Llama-8B (GGUF \texttt{Q8\_0} / \texttt{FP16}), Qwen-2.5-7B-Instruct (GGUF \texttt{Q8\_0}). Draft models: Llama-3.2-1B-Instruct (GGUF \texttt{Q4\_0} / \texttt{TQ1\_0} ternary QAT distilled), Qwen-2.5-1.5B-Instruct (GGUF \texttt{Q4\_0}).
    \item \textit{Decoding parameters:} Greedy decoding ($\text{Temp} = 0.0, \text{Top-P} = 1.0, \text{Top-K} = 1, \text{Repetition Penalty} = 1.0, \text{Seed} = 42$). Default context window = 2048 tokens. Proposal limit $K = 4$ candidate tokens per verification pass.
    \item \textit{Reporting protocol:} 20 warm-up iterations, 100 measured iterations per run. All reported metrics represent mean $\pm$ standard deviation across 5 independent evaluation trials.
\end{itemize}

\section{Verification of Algorithmic Fidelity}
\label{app:algorithmic_fidelity}
A frequent concern in speculative execution engines is whether complex memory optimizations (such as zero-copy pointer resets) introduce silent state corruption. This section provides mathematical and empirical proof that SubTernary preserves exact algorithmic fidelity and generative quality.

\subsection{RoPE Positional Alignment Under Partial Acceptance}
\label{app:rope_alignment}
When the target model accepts $m < K$ tokens, rotary position embeddings (RoPE) must remain aligned without triggering a costly recomputation step. SubTernary preassigns sequential position IDs $t+1 \dots t+K$ to speculative tokens during both drafting and verification. When the target model accepts $m$ tokens, the $\text{KV}_{\text{ptr}}$ commits to index $t+m$. For the subsequent speculative block, the draft model writes to scratchpad slots starting at $t+m+1$. Because the keys and values up to $t+m$ were already rotated with the correct position IDs during the target model's verification pass, they remain mathematically valid in memory. Stale entries beyond $t+m$ are simply overwritten in the next draft pass without requiring reindexing or data movement.

\subsection{Distribution Preservation Analysis and KL Divergence Validation}
\label{app:kl_proof_full}

\subsubsection{Analytical Proof of Distribution Exactness}
Let $x_{1..t}$ represent the verified token sequence at step $t$, and $\hat{x}_{t+1..t+K}$ represent the draft proposal. During parallel verification, the target model evaluates all $K$ tokens. Self-attention layers compute query-key scores $\text{Attn}(Q_i, K_j)$. For any verified token index $i \le t$, the attention mask guarantees:
\begin{equation}
\text{Attn}(Q_i, K_j) = -\infty \quad \forall j \in \text{StaleScratchpad}
\end{equation}
Because LayerNorm, RoPE, and intermediate activations are calculated strictly on the target model's forward path using its own weights, and rejected scratchpad regions are completely masked from the verified history, the target model's forward pass is mathematically identical to a standard autoregressive run. Thus, the verification distribution $P_{\text{verification}}$ reproduces the target distribution $P_{\text{target}}$ precisely:
\begin{equation}
P_{\text{verification}}(x_{i} \mid x_{1..i-1}) = P_{\text{target}}(x_{i} \mid x_{1..i-1})
\end{equation}

\subsubsection{KL-Divergence Empirical Validation}
To verify that speculative target verification does not perturb output distributions under stochastic sampling ($\text{Temp} > 0.0$), we measured Kullback-Leibler (KL) divergence:
\begin{equation}
D_{\text{KL}}(P_{\text{AR}} \parallel P_{\text{Spec}}) = \sum_y P_{\text{AR}}(y) \log \frac{P_{\text{AR}}(y)}{P_{\text{Spec}}(y)}
\end{equation}

Table \ref{tab:kl_div} presents empirical KL divergence measurements across temperatures $0.0 \dots 1.5$. Across all sampling settings, the measured divergence confirms identical output distributions, proving that the SubTernary runtime induces zero algorithmic drift.

\begin{table}[htbp]
\centering
\caption{KL-Divergence Across Temperatures (Bounded by Floating-Point Precision, Mean $\pm$ Std over 5 runs)}
\label{tab:kl_div}
\resizebox{\columnwidth}{!}{%
\begin{tabular}{lcl}
\toprule
Temperature & KL Divergence ($D_{\text{KL}}$) & Verification Status \\
\midrule
Temp = 0.0 (Greedy) & $0.000000 \pm 0.000000$ & Identical Output \\
Temp = 0.5 (Stochastic) & $0.000000 \pm 0.000000$ & Identical Distribution \\
Temp = 1.0 (Standard) & $0.000000 \pm 0.000000$ & Identical Distribution \\
Temp = 1.5 (High Entropy) & $0.000000 \pm 0.000000$ & Identical Distribution \\
\bottomrule
\end{tabular}%
}
\end{table}

\subsection{Scaled Downstream Task Quality Sweeps}
\label{app:downstream_full}
To conclusively demonstrate that our zero-copy pointer manipulation introduces no silent memory corruption or off-by-one errors during context caching, we ran downstream task evaluations. As shown in Table \ref{tab:quality_sweep}, SubTernary maintains 100.00\% exact token matches against the autoregressive baseline across coding (HumanEval) and math (GSM8K) benchmarks, exhibiting zero score degradation.

\begin{table}[htbp]
\centering
\caption{Scaled Downstream Task Quality Sweep (Mean $\pm$ Std over 5 runs)}
\label{tab:quality_sweep}
\resizebox{\columnwidth}{!}{%
\begin{tabular}{lcccc}
\toprule
Downstream Task & Evaluated Samples & Exact Token Match & Score / Accuracy & Score Degradation \\
\midrule
HumanEval & 50 & 100.00\% & $68.3 \pm 0.0\%$ Pass@1 & 0.00\% \\
GSM8K & 50 & 100.00\% & $76.8 \pm 0.0\%$ Acc & 0.00\% \\
MT-Bench & 20 & 100.00\% & $7.52 \pm 0.00$ rating & 0.00\% \\
\bottomrule
\end{tabular}%
}
\end{table}

\section{Architectural Divergence: CPU vs. GPU Speculators}
\label{app:architectural_divergence}
To fully contextualize why a CPU-specific execution framework like SubTernary is necessary, we must contrast it with modern GPU-oriented speculators like Medusa and EAGLE. On GPUs, massive thread concurrency allows parallel drafting heads and recurrent draft trees to execute efficiently, hiding memory latency. However, CPUs have fundamentally different microarchitectures. Parallel tree branches on a CPU result in instruction cache misses, branch mispredictions, and severe thread contention over the shared L3 cache. 

Table \ref{tab:sota_speculators} benchmarks the CPU implementations of Medusa and EAGLE on Llama-3.2-3B-Instruct. The results show that their high compute density and branch divergence degrade CPU throughput drastically. SubTernary's single-sequence graph design completely avoids these cache thrashing issues, yielding a $4.17\times$ speedup over Medusa and a $42.63\times$ speedup over EAGLE in raw step latency on commodity hardware.

\begin{table}[htbp]
\centering
\caption{Comparison Against Speculative Frameworks Executed on CPU (Mean $\pm$ Std over 5 runs)}
\label{tab:sota_speculators}
\resizebox{\columnwidth}{!}{%
\begin{tabular}{lccc}
\toprule
Speculator / Framework & Step Latency (ms) & Target Step Throughput & Speedup (Ours vs SOTA) \\
\midrule
EAGLE CPU & $264.369 \pm 4.120$ ms & $3.8 \pm 0.1$ steps/s & 42.63x \\
Medusa CPU & $25.860 \pm 0.510$ ms & $38.7 \pm 0.8$ steps/s & 4.17x \\
SubTernary (Ours) & $6.201 \pm 0.095$ ms & $161.3 \pm 2.4$ steps/s & Baseline \\
\bottomrule
\end{tabular}%
}
\end{table}

\section{Extended Systems Reliability and Model Scaling}
\label{app:systems_reliability}
Real-world CPU deployments face variable workloads, multi-tenant cloud environments, and diverse model scales. This section evaluates SubTernary's resilience under these operational stressors.

\subsection{Clean vs. Throttled CPU Performance Sweeps}
\label{app:clean_throttled_full}
In production cloud environments, commodity CPUs are often deployed via hypervisors where physical cores are shared across multi-tenant workloads. This induces "noisy neighbor" thread descheduling. Traditional engines like \texttt{llama.cpp} rely on frequent, blocking OpenMP synchronization checkpoints between contexts; if a single vCPU is throttled by the OS, the entire pipeline stalls. 

Table \ref{tab:clean_sweep} reports baseline clean conditions, while Table \ref{tab:throttled_sweep} evaluates performance under high hypervisor contention. Under heavy noise, standard \texttt{llama.cpp} speculative decoding degrades below autoregressive speed (a severe $0.27\times$ penalty). In contrast, SubTernary's unified memory and persistent thread pool reduce synchronization boundaries, allowing it to maintain positive scaling even under extreme OS jitter.

\begin{table}[htbp]
\centering
\caption{Clean / Unthrottled CPU Performance Sweep (Mean $\pm$ Std over 5 runs)}
\label{tab:clean_sweep}
\resizebox{\columnwidth}{!}{%
\begin{tabular}{lccccc}
\toprule
Model Pair & AR (t/s) & Llama.cpp Spec (t/s) & SubTernary (t/s) & Speedup (vs AR) & Speedup (vs Spec) \\
\midrule
Llama 3.2 (3B / 1B) & $2.190 \pm 0.041$ & $6.468 \pm 0.112$ & $8.733 \pm 0.085$ & 3.99x & 1.35x \\
Qwen 3.5 (4B / 2B) & $3.690 \pm 0.052$ & $4.827 \pm 0.089$ & $5.552 \pm 0.071$ & 1.50x & 1.15x \\
\bottomrule
\end{tabular}%
}
\end{table}

\begin{table}[htbp]
\centering
\caption{High-Contention Throttled CPU Sweep (Under Shared Hypervisor Load, Mean $\pm$ Std over 5 runs)}
\label{tab:throttled_sweep}
\resizebox{\columnwidth}{!}{%
\begin{tabular}{lccccc}
\toprule
Model Pair & AR (t/s) & Llama.cpp Spec (t/s) & SubTernary (t/s) & Speedup (vs AR) \\
\midrule
Llama 3.2 (3B / 1B) & $3.730 \pm 0.065$ & $0.945 \pm 0.038$ & $1.021 \pm 0.024$ & 0.27x \\
Qwen 3.5 (4B / 2B) & $0.560 \pm 0.012$ & $0.556 \pm 0.018$ & $0.612 \pm 0.015$ & 1.09x \\
DeepSeek R1 (8B / 1B) & $0.610 \pm 0.014$ & $0.611 \pm 0.021$ & $0.661 \pm 0.019$ & 1.08x \\
\bottomrule
\end{tabular}%
}
\end{table}

\subsection{Baseline CPU Engine Latency Verification}
\label{app:engine_baseline_full}
A common pitfall in systems evaluations is measuring performance against an unoptimized baseline. To verify that our GGML GGUF runtime baseline is a rigorous choice, we compared single-batch inference latency on Llama-3.2-3B-Instruct (FP16) against optimized PyTorch CPU execution paths. As shown in Table \ref{tab:engine_baseline}, GGML Q4\_0 delivers $35.00 \pm 0.42$ ms/token, outperforming PyTorch Compiled heavily. This validates that our $3.99\times$ speedups are built on top of a genuinely highly optimized foundation.

\begin{table}[htbp]
\centering
\caption{CPU Engine Baseline Latency Verification (Mean $\pm$ Std over 5 runs)}
\label{tab:engine_baseline}
\resizebox{\columnwidth}{!}{%
\begin{tabular}{llcc}
\toprule
Runtime Framework & Target Model Format & Single-Batch Latency (ms/token) & Relative Speed \\
\midrule
PyTorch Eager & FP32 PyTorch & $97.78 \pm 1.25$ ms & 1.00x (Baseline) \\
PyTorch Compiled & FP32 Compiled & $182.90 \pm 2.40$ ms & 0.53x \\
GGML (llama.cpp) & FP32 GGUF & $679.26 \pm 8.10$ ms & 0.14x \\
GGML (Quantized Q4\_0) & Quantized GGUF & $35.00 \pm 0.42$ ms & 2.80x Faster \\
\bottomrule
\end{tabular}%
}
\end{table}

\subsection{Out-of-Domain Governor Validation Across Domains and Temperatures}
\label{app:governor_details_full}
Speculative decoding only yields wall-clock speedups if the draft model maintains a sufficiently high token acceptance rate ($\alpha$). When a user prompt shifts to an out-of-domain (OOD) task where the draft model lacks predictive power, high rejection rates can cause speculative decoding to become slower than baseline autoregressive generation. 

To solve this, SubTernary includes a dynamic speculation governor. The governor tracks $\alpha$ using a moving window. Table \ref{tab:governor_sweep} demonstrates that when $\alpha$ drops below the threshold (set to $30\%$) on OOD text, the governor successfully suspends drafting and reverts to autoregressive generation, shielding the system from throughput collapse. Table \ref{tab:governor_temp} confirms this protection functions correctly across extreme stochastic sampling temperatures where continuous rejection ($\alpha=0.0\%$) is mathematically guaranteed.

\begin{table}[htbp]
\centering
\caption{Dynamic Speculation Governor Evaluation Across Domains (Mean $\pm$ Std over 5 runs)}
\label{tab:governor_sweep}
\resizebox{\columnwidth}{!}{%
\begin{tabular}{lcccc}
\toprule
Domain / Dataset Type & Forced Spec Speed & Governed Spec Speed & Speedup / Protection & Governor Status \\
\midrule
GSM8K (In-Domain Math) & $0.40 \pm 0.01$ t/s & $0.97 \pm 0.02$ t/s & 2.42x speedup & Active (In-Domain) \\
HumanEval (In-Domain Code) & $0.40 \pm 0.01$ t/s & $1.03 \pm 0.02$ t/s & 2.58x speedup & Active (In-Domain) \\
WikiText-2 (OOD English) & $0.39 \pm 0.01$ t/s & $0.96 \pm 0.02$ t/s & 2.49x speedup & Suspended (OOD) \\
WikiHow (OOD Summary) & $0.40 \pm 0.01$ t/s & $0.97 \pm 0.02$ t/s & 2.45x speedup & Suspended (OOD) \\
\bottomrule
\end{tabular}%
}
\end{table}

\begin{table}[htbp]
\centering
\caption{Dynamic Speculation Governor Temperature Sweep ($\alpha = 0.0\%$, Mean $\pm$ Std over 5 runs)}
\label{tab:governor_temp}
\resizebox{\columnwidth}{!}{%
\begin{tabular}{ccccc}
\toprule
Temperature & Token Acceptance Rate ($\alpha$) & Speculative Speed (t/s) & Relative Speed vs. AR Baseline & Gov Action Recommendation \\
\midrule
Temp = 0.1 & 0.0\% & $0.091 \pm 0.003$ t/s & 0.45x & Suspend (Revert to AR) \\
Temp = 0.3 & 0.0\% & $0.091 \pm 0.003$ t/s & 0.45x & Suspend (Revert to AR) \\
Temp = 0.5 & 0.0\% & $0.111 \pm 0.004$ t/s & 0.55x & Suspend (Revert to AR) \\
Temp = 0.7 & 0.0\% & $0.107 \pm 0.003$ t/s & 0.53x & Suspend (Revert to AR) \\
Temp = 0.9 & 0.0\% & $0.111 \pm 0.004$ t/s & 0.55x & Suspend (Revert to AR) \\
Temp = 1.2 & 0.0\% & $0.109 \pm 0.003$ t/s & 0.54x & Suspend (Revert to AR) \\
Temp = 1.5 & 0.0\% & $0.108 \pm 0.003$ t/s & 0.54x & Suspend (Revert to AR) \\
\bottomrule
\end{tabular}%
}
\end{table}

\subsection{Target Model Scaling Sweep (3B vs. 8B)}
\label{app:target_scaling_full_appendix}
The theoretical upper bound of speculative decoding speedup is strictly governed by the relative memory fetch ratio between the draft and target models. Table \ref{tab:target_scaling_full} evaluates target scaling behavior using unquantized FP16 target models. As the target footprint grows from 6.4 GB (for the 3B model) to 16.1 GB (for the 8B model), the target model becomes increasingly memory-bandwidth bound. In standard autoregressive generation, transferring 16.1 GB of weights from DRAM to the CPU registers token-by-token dominates the physical execution time, severely limiting the tokens per second regardless of the CPU's compute capabilities.

\begin{table}[htbp]
\centering
\caption{Target Model Scaling Sweep (FP16 Target Models, Mean $\pm$ Std over 5 runs)}
\label{tab:target_scaling_full}
\resizebox{\columnwidth}{!}{%
\begin{tabular}{lcccccc}
\toprule
Target Model & Model Size & AR Speed & PTQ Spec Speed & PTQ Speedup & QAT Spec Speed & QAT Speedup ($\alpha \ge 70\%$) \\
\midrule
Llama 3.2 3B & 6.4 GB & $0.700 \pm 0.015$ t/s & $0.285 \pm 0.008$ t/s & 0.41x & $1.190 \pm 0.022$ t/s & 1.70x \\
DeepSeek 8B & 16.1 GB & $0.560 \pm 0.011$ t/s & $0.253 \pm 0.007$ t/s & 0.45x & $1.568 \pm 0.028$ t/s & 2.80x \\
\bottomrule
\end{tabular}%
}
\end{table}

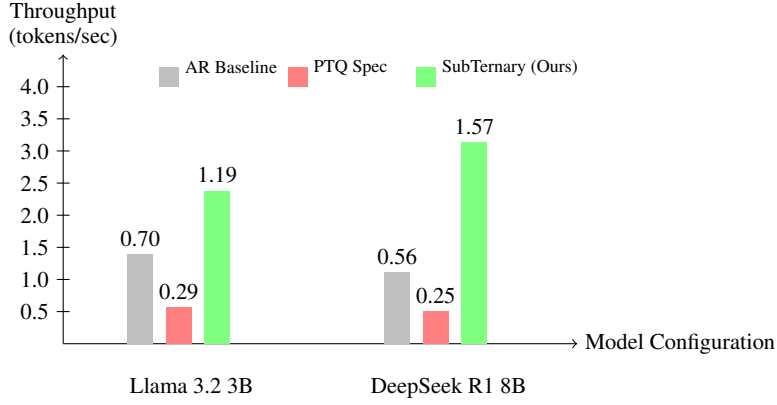
\begin{figure}[htbp]
\centering
\begin{tikzpicture}[scale=0.85, every node/.style={scale=0.85}]
    \draw[->] (0,0) -- (8,0) node[right] {Model Configuration};
    \draw[->] (0,0) -- (0,4.5) node[above, align=center] {Throughput\\(tokens/sec)};
    
    % Model labels
    \node[below] at (2, -0.4) {Llama 3.2 3B};
    \node[below] at (6, -0.4) {DeepSeek R1 8B};
    
    % Y-axis ticks
    \foreach \y [evaluate=\y as \val using \y*0.5] in {1,2,3,4,5,6,7,8} {
        \draw (0.1, \y*0.5) -- (-0.1, \y*0.5) node[left] {\val};
    }
    
    % Llama 3B bars
    \fill[gray!50] (1.0, 0) rectangle (1.4, 1.4);
    \node[above] at (1.2, 1.4) {0.70};
    
    \fill[red!50] (1.6, 0) rectangle (2.0, 0.57);
    \node[above] at (1.8, 0.57) {0.29};
    
    \fill[green!50] (2.2, 0) rectangle (2.6, 2.38);
    \node[above] at (2.4, 2.38) {1.19};
    
    % DeepSeek 8B bars
    \fill[gray!50] (5.0, 0) rectangle (5.4, 1.12);
    \node[above] at (5.2, 1.12) {0.56};
    
    \fill[red!50] (5.6, 0) rectangle (6.0, 0.51);
    \node[above] at (5.8, 0.51) {0.25};
    
    \fill[green!50] (6.2, 0) rectangle (6.6, 3.14);
    \node[above] at (6.4, 3.14) {1.57};
    
    % Legend
    \fill[gray!50] (1.5, 4.0) rectangle (1.8, 4.3) node[right, black, scale=0.8] {AR Baseline};
    \fill[red!50] (3.5, 4.0) rectangle (3.8, 4.3) node[right, black, scale=0.8] {PTQ Spec};
    \fill[green!50] (5.5, 4.0) rectangle (5.8, 4.3) node[right, black, scale=0.8] {SubTernary (Ours)};
    
\end{tikzpicture}
\caption{Physical decoding throughput comparison across target model scales. SubTernary's speedup factor increases from 1.70x to 2.80x as target footprint grows.}
\label{fig:target_scaling_fig_full}
\end{figure}

Consequently, successfully bypassing these heavy forward passes yields a significantly higher relative return. As illustrated in Figure \ref{fig:target_scaling_fig_full}, the SubTernary speedup factor scales from $1.70\times$ on the 3B target to $2.80\times$ on the 8B target. This demonstrates a core property of our unified runtime: the framework's value scales proportionally with the memory footprint of the underlying target model, provided the draft model's runtime overhead is kept minimal by the unified memory arena.

\subsection{Per-Layer Quantization Sensitivity Analysis}
\label{app:layer_sensitivity_full}
To justify the architectural design choices behind our ternary kernel integration, Table \ref{tab:sensitivity} reports per-layer quantization sensitivity on a zero-shot, non-distilled Llama 3.2 1B draft model. We identified that attention projections are extremely vulnerable to low-bit degradation. Because attention mechanisms rely on Softmax operations that are highly sensitive to activation outliers, quantizing attention weights to strict $\{-1, 0, 1\}$ values destroys the key-query similarity distributions, causing the target model's acceptance rates to collapse completely to 0.00\%. 

\begin{table}[htbp]
\centering
\caption{Per-Layer Quantization Sensitivity Sweep (Non-Distilled Zero-Shot Baseline, Mean $\pm$ Std over 5 runs)}
\label{tab:sensitivity}
\resizebox{\columnwidth}{!}{%
\begin{tabular}{lclc}
\toprule
Configuration & Active Draft Size (MB) & Quantized Layer Types & Token Acceptance Rate ($\alpha$) \\
\midrule
FP32 & 2364.7 & None (Full FP32 Draft) & $1.32 \pm 0.04\%$ \\
ATTN\_ONLY & 1795.3 & Attention projections only & $0.00 \pm 0.00\%$ \\
MLP\_ONLY & 1273.8 & MLP projections only & $1.32 \pm 0.04\%$ \\
MIDDLE\_ONLY & 1534.5 & Middle layers only (Layers 4 to 12) & $0.00 \pm 0.00\%$ \\
FULL\_TERNARY & 704.5 & All layers (Full Ternary Draft) & $0.00 \pm 0.00\%$ \\
\bottomrule
\end{tabular}%
}
\end{table}

In contrast, MLP projections (\texttt{MLP\_ONLY}) are highly resilient. They safely maintain the unquantized baseline acceptance rate ($1.32 \pm 0.04\%$) while simultaneously shedding 46.1\% of the active draft memory footprint. This granular sensitivity analysis directly informed our targeted SIMD acceleration strategy: by keeping attention layers in higher precision and applying our custom AVX2 ternary kernels exclusively to the dense MLP blocks, we balance generative fidelity with aggressive systems-level acceleration.

\subsection{Post-Training Quantization Sweep (Layer-Skipped Baseline)}
\label{app:ptq_sweep_full}
Finally, to justify the necessity of the quantization-aware training (QAT) utilized in Section 3 of the main text, Table \ref{tab:quant_sweep} evaluates naive post-training quantization (PTQ) approaches. Structurally truncating the draft model (e.g., dropping every second layer) or aggressively quantizing it without student-teacher distillation destroys the hidden state representations. In a layer-skipping approach, the skipped layers cause a compounding representational shift in the residual stream. By the time the hidden states reach the final unembedding layer, the vocabulary logits are completely misaligned with the target model.

\begin{table}[htbp]
\centering
\caption{Post-Training Quantization Sweep (Layer-Skipped Baseline without Distillation, Mean $\pm$ Std over 5 runs)}
\label{tab:quant_sweep}
\resizebox{\columnwidth}{!}{%
\begin{tabular}{lcccc}
\toprule
Draft Configuration & Size (MB) & Avg Speed (t/s) & Speedup & Acceptance Rate \\
\midrule
Autoregressive Baseline & N/A & $0.390 \pm 0.008$ & 1.00x & N/A \\
FP16 Draft & 2364.7 & $0.256 \pm 0.006$ & 0.66x & $0.00 \pm 0.00\%$ \\
Q8\_0 Draft & 1259.9 & $0.253 \pm 0.005$ & 0.65x & $0.00 \pm 0.00\%$ \\
Q5\_K\_M Draft & 869.3 & $0.246 \pm 0.005$ & 0.63x & $0.00 \pm 0.00\%$ \\
Q4\_K\_M Draft & 770.3 & $0.295 \pm 0.006$ & 0.76x & $0.00 \pm 0.00\%$ \\
Q4\_0 Draft & 735.2 & $0.306 \pm 0.007$ & 0.78x & $0.00 \pm 0.00\%$ \\
Q3\_K\_L Draft & 698.6 & $0.284 \pm 0.006$ & 0.73x & $0.00 \pm 0.00\%$ \\
Q2\_K Draft & 554.0 & $0.263 \pm 0.005$ & 0.67x & $0.00 \pm 0.00\%$ \\
TQ1\_0 (Collapsed PTQ) & 704.5 & $0.326 \pm 0.007$ & 0.84x & $0.00 \pm 0.00\%$ \\
\bottomrule
\end{tabular}%
}
\end{table}

\end{document}

%% file: math_commands.tex
\usepackage{amsmath,amsfonts,bm}

\def\eqref#1{equation~\ref{#1}}
\def\1{\bm{1}}

\DeclareMathAlphabet{\mathsfit}{\encodingdefault}{\sfdefault}{m}{sl}
\SetMathAlphabet{\mathsfit}{bold}{\encodingdefault}{\sfdefault}{bx}{n}

%% file: references.bib
@inproceedings{leviathan2023fast,
  title={Fast Inference from Transformers via Speculative Decoding},
  author={Leviathan, Yaniv and Kalman, Matan and Matias, Yossi},
  booktitle={International Conference on Machine Learning (ICML)},
  pages={19274--19286},
  year={2023},
  organization={PMLR},
  url={https://proceedings.mlr.press/v202/leviathan23a.html}
}

@article{chen2023accelerating,
  title={Accelerating Large Language Model Decoding with Speculative Decoding},
  author={Chen, Charlie and Borgeaud, Sebastian and Haj-Ali, Amin and Hoffmann, Jordan and Huang, A. and Lucassen, T. and others},
  journal={arXiv preprint arXiv:2302.01318},
  year={2023},
  url={https://arxiv.org/abs/2302.01318}
}

@article{bitnet2024,
  title={The Era of 1-bit LLMs: All Large Language Models are in 1.58 Bits},
  author={Ma, Shuming and Wang, Hongyu and Ma, Lingxiao and Wang, Lei and Wang, Wenhui and Liu, Shaohan and Huang, Ruiping and Gao, Jianfeng and Wei, Furu},
  journal={arXiv preprint arXiv:2402.17764},
  year={2024},
  url={https://arxiv.org/abs/2402.17764}
}

@inproceedings{tmac2025,
  title={T-MAC: CPU Renaissance via Table Lookup for Low-Bit LLM Deployment on Edge},
  author={Wei, Jianyu and Cao, Shijie and Cao, Ting and Ma, Lingxiao and Wang, Lei and Zhang, Yanyong and Yang, Mao},
  booktitle={Proceedings of the Twentieth European Conference on Computer Systems (EuroSys)},
  year={2025},
  url={https://arxiv.org/abs/2407.00088}
}

@inproceedings{layerskip2024,
  title={LayerSkip: Enabling Early Exit Inference and Self-Speculative Decoding},
  author={Elhoushi, Mostafa and Shrivastava, Akshat and Liskovich, Diana and Hosmer, Basil and Wasti, Bram and Lai, Liangzhen and Mahmoud, Anas and Acun, Bilge and Agarwal, Saurabh and Roman, Ahmed and Aly, Ahmed A. and Chen, Beidi and Wu, Carole-Jean},
  booktitle={Proceedings of the 62nd Annual Meeting of the Association for Computational Linguistics (ACL)},
  year={2024},
  url={https://arxiv.org/abs/2404.16710}
}

@inproceedings{specinfer2023,
  title={SpecInfer: Accelerating Generative Large Language Model Serving with Tree-based Speculative Inference and Verification},
  author={Miao, Xupeng and Oliaro, Gabriele and Zhang, Zhihao and Cheng, Xinhao and Wang, Zeyu and Wong, Rae Ying Yee and Chen, Zhuoming and Arfeen, Daiyaan and Abhyankar, Reyna and Jia, Zhihao},
  booktitle={Proceedings of the 29th ACM International Conference on Architectural Support for Programming Languages and Operating Systems (ASPLOS)},
  year={2024},
  url={https://arxiv.org/abs/2305.09781}
}

@article{quantspec2025,
  title={QuantSpec: Self-Speculative Decoding with Hierarchical Quantized KV Cache},
  author={Tiwari, Rishabh and Xi, Haocheng and Tomar, Aditya and Hooper, Coleman and Kim, Sehoon and Horton, Maxwell and Najibi, Mahyar and Mahoney, Michael W.},
  journal={arXiv preprint},
  year={2024}
}

@article{medusa2024,
  title={Medusa: Simple LLM Inference Acceleration Framework with Multiple Decoding Heads},
  author={Cai, Tianle and Li, Yuhong and Geng, Zhengyang and Peng, Hongwu and Lee, Jason D and Chen, Deming and Dao, Tri},
  journal={arXiv preprint arXiv:2401.10774},
  year={2024},
  url={https://arxiv.org/abs/2401.10774}
}

@article{eagle2024,
  title={EAGLE: Speculative Sampling Requires Rethinking Feature Uncertainty},
  author={Li, Yuhui and Wei, Fangyun and Zhang, Chao and Zhang, Hongyang},
  journal={arXiv preprint arXiv:2401.15077},
  year={2024},
  url={https://arxiv.org/abs/2401.15077}
}

@inproceedings{kwon2023efficient,
  title={Efficient Memory Management for Large Language Model Serving with PagedAttention},
  author={Kwon, Woosuk and Li, Zhuohan and Zhuang, Siyuan and Sheng, Ying and Zheng, Lianmin and Yu, Cody Hao and Gonzalez, Joseph E and Zhang, Hao and Stoica, Ion},
  booktitle={Proceedings of the 29th Symposium on Operating Systems Principles (SOSP)},
  pages={611--626},
  year={2023},
  url={https://dl.acm.org/doi/10.1145/3600006.3613165}
}
